\documentclass{ws-ijmpd}

\usepackage{amssymb}

\begin{document}

\markboth{F.Cannata and A.Yu.Kamenshchik}
{Chameleon cosmology model describing the phantom divide line crossing}

%
\catchline{}{}{}{}{}
%

\title{CHAMELEON COSMOLOGY MODEL DESCRIBING THE PHANTOM DIVIDE LINE CROSSING}


\author{FRANCESCO CANNATA}
\address{INFN, sezione di Bologna, Via Irnerio 46, 40126 Bologna, Italy\\
cannata@bo.infn.it}
\author{ALEXANDER Yu KAMENSHCHIK}
\address{
Dipartimento di Fisica and INFN, sezione di Bologna, Via Irnerio 46, 40126 Bologna,Italy\\
L.D. Landau Institute for Theoretical Physics of the Russian Academy of Sciences, Kosygin str.~2, 119334 Moscow,
Russia\\ 
kamenshchik@bo.infn.it}

\maketitle

\begin{abstract}
An exact solution describing the evolution of the type Bang-to-Rip with the phantom divide line crossing is constructed 
in the Chameleon cosmology model, based on two independent functions of the scalar field.   
\end{abstract}

\keywords{dark energy; chameleon cosmology; phantom divide line crossing}

\section{Introduction}
The discovery of cosmic acceleration\cite{cosmic} has stimulated an intensive study of  models
of dark energy\cite{dark} responsible for the origin of this phenomenon. Dark energy is characterized by a
negative pressure whose relation to the energy density $w=p/\varepsilon$ is less than $-1/3$. Moreover, if this
relation happens to be less than $-1$ such kind of dark energy is called ``phantom'' dark energy\cite{phantom}.
The breakdown of the condition $w > -1$  implies the so
called super-acceleration of cosmological evolution, which in some models culminates approaching a new type of
cosmological singularity called Big Rip\cite{Rip}. While the cosmological constant ($w = -1$) is still a
possible candidate for the role of dark energy, there are observations giving  indications in favor of the models
where the equation of state parameter $w$ not only changes with time, 
but could become less than $-1$ no\cite{observe}. 
There are also observations \cite{obs-cross} indicating the phenomenon of the phantom divide line crossing \cite{divide},
when the parameter $w$ changes between the phantom and non-phantom phases.  
A standard way to introduce the phantom energy is to consider a scalar  field with the negative
sign of the kinetic energy term, usually called ``phantom'' scalar field. The simplest way of describing the phenomenon 
of the phantom divide line crossing is to introduce two scalar fields - phantom and non-phantom as in the so called quintom models 
\cite{two-exp}.  Some other models are based on the use of the PT symmetry ideas \cite{we-PT,we-PT1,we-PT2}. 
Note that the phantom divide line crossing can occur also in models based on one scalar field if some special potential are chosen 
\cite{we} or if the non-minimal coupling between a scalar field and gravity is included \cite{non-minimal}.

Naturally, consideration of two scalar fields open wider opportunities for description of any prescribed cosmological evolution, than 
working with a single field \cite{we-two}.

Recently, another class of cosmological dark energy models was put forward - the so called chameleon cosmology models \cite{chameleon}.
In this models the scalar field interacts with some kind of matter, behaving as perfect fluid. The Lagrangian of the model contains 
a term, where the matter interacts with the effective metric (physical metric, multiplied by a 
conformal factor depending on the scalar field). The energy density of the perfect fluid is proportional 
to the trace of its energy-momentum tensor, multiplied by this conformal factor. On the other hand the effective 
mass of the scalar field becomes variable, depending on the energy density of matter, which changes in the process 
of the cosmological evolution.  

Lately, an alternative  version of the Chameleon cosmology was proposed in paper \cite{Fara-Sale}. 
Here the term, describing the interaction between the chameleon scalar field and matter has a very general 
multiplicative form and is equal to some arbitrary function of the scalar field multiplied by the Lagrangian 
of  matter, which on the Friedmann-Robertson-Walker background coincides with the pressure of the matter. 
Thus, one has two arbitrary functions of the scalar field to play with - the standard potential and the potential-like 
function describing the interaction of matter with the chameleon field. Choosing properly  these two functions of the scalar 
field one can describe different kinds of cosmological evolution, including those where the universe undergoes the 
phantom divide line crossing. The authors of paper \cite{Fara-Sale} use the numerical integration of the corresponding 
equations of motion and compare the results with the data coming from time evolution of the cosmological redshift of 
distant sources. 

We remark that having two functions of a scalar field one gets a sufficient freedom to construct exactly solvable solutions 
of the cosmological equations, giving some prescribed cosmological evolution. Just like choosing exponential potentials
one can find exact solutions desribing power-law expansion of the universe (see e.g. \cite{power-law}), here one can try 
to construct two potential-like functions providing the existence of exact cosmological solutions, undergoing the phantom 
divide line crossing. In fact, this is the task of the present paper.

We consider the flat Friedmann model with the metric 
\begin{equation}
ds^2 = dt^2 - a^2(t)dl^2.
\label{Friedmann}
\end{equation}
The Hubble parameter behaves as 
\begin{equation}
H(t) \equiv \frac{\dot{a}(t)}{a(t)} = \frac{\alpha t_{R}}{t(t_R-t)},
\label{Hubble}
\end{equation}
where $\alpha$ is some positive parameter of the order $1$ and ``dot'' means as usual the derivative with respect to 
the cosmic time $t$.
Such a cosmological evolution was considered in papers \cite{we-PT1,we-two} in the the context of two-field models 
and was called ``Bang-to-Rip'' evolution. Here the universe begins its evolution from the Big Bang singularity 
when the Hubble parameter behaves as $H(t) \sim \frac{\alpha}{t}$ and ends in the Big Rip singularity at the moment 
$t = t_R$. At the moment $t = \frac{t_R}{2}$ the phantom divide line is crossed by the universe. 
Indeed, 
the phantom divide line crossing means that the relation between the pressure and the energy density becomes equal to $-1$.
That means that in the energy conservation law
$\dot{\varepsilon} = -3H(\varepsilon+p)$
the right-hand side vanishes, and so does the time derivative of the total energy density in its left-hand side.  
 The energy density is proportinal in the flat Friedmann model to the Hubble parameter squared.  
Hence, the time derivative of the Hubble parameter $H$ should vanish too.
The time derivative of the Hubble parameter from Eq. (\ref{Hubble}) is
\begin{equation}
\dot{H} = \frac{2\alpha t_R(2t-t_R)}{t^2(t_R-t)^2}
\label{Hubble1}
\end{equation}
and vanishes at $t = t_R/2$.

We shall consider the  minimally coupled to gravity scalar field $\phi$ with the potential $V(\phi)$ whose interaction 
with the perfect fluid is described by the term in the  Lagrangian, which on the Friedmann background looks like \cite{Fara-Sale}:    
\begin{equation}
L_{scalar+matter} = \gamma\rho f(\phi),
\label{Cham}
\end{equation}
where the coefficient $\gamma$ relates the energy density $\rho$ and the pressure $p$ of matter:
\begin{equation}
p = \gamma \rho,
\label{eq-of-state}
\end{equation}
and $f(\phi)$ is some function of the scalar field $\phi$. 
In the next section we shall find the form of the functions $V(\phi)$ and $f(\phi)$, and the exact solution 
for $a(t)$ and $\phi(t)$ matching the evolution (\ref{Hubble}). 
In the third section we discuss the bounds on the parameters of the model.
The last section contains some concluding remarks.  

\section{An exactly solvable Chameleon field cosmological model}
We shall fix the fundamental constants in such a way to give to the Friedmann equation a particularly simple form:
\begin{equation}
H^2 = \varepsilon,
\label{Friedmann1}
\end{equation}
where $\varepsilon$ is the total energy density of the scalar field and matter. On the flat Friedmann background 
this total energy density is 
\begin{equation}
\varepsilon = \frac{\dot{\phi}^2}{2} + V(\phi) + \rho f(\phi).
\label{energy}
\end{equation}
The Klein-Gordon equation for the scalar field $\phi$ is 
\begin{equation}
\ddot{\phi} + 3H\dot{\phi} + V'(\phi) + \gamma \rho f'(\phi) = 0,
\label{KG}
\end{equation}
 where ``prime'' stays for the derivative with respect to $\phi$. 
The total energy density $\varepsilon$ satisfies the energy conservation law 
\begin{equation}
\dot{\varepsilon} + 3H(\varepsilon + P) = 0,
\label{conserve}
\end{equation}
where $P$ is the total pressure of the matter and of the scalar field which is equal to 
\begin{equation}
P = \gamma \rho + \frac{\dot{\phi}^2}{2} - V(\phi).
\label{pressure}
\end{equation}
Substituting Eqs. (\ref{energy}) and (\ref{pressure}) into Eq. (\ref{conserve}) we obtain
\begin{equation}
\dot{\phi} (\ddot{\phi} + 3H\dot{\phi} + V' + \rho f') 
=-3 (1+\gamma) H \rho f -\dot{\rho} f.
\label{conserve1}
\end{equation}

Using Eq. (\ref{KG}) we can simplify the left-hand side of Eq. (\ref{conserve1}) coming to 
\begin{equation}
(1-\gamma) \rho \dot{f} + 3 (1+\gamma) H \rho f + \dot{\rho} f = 0.
\label{conserve2}
\end{equation}
Integrating Eq. (\ref{conserve2}) one obtains for the energy density of the matter 
\begin{equation}
\rho(t) = \frac{\rho_0}{f^{1-\gamma}(\phi(t)) a^{3(1+\gamma)}(t)},
\label{matter}
\end{equation}
where $\rho_0$ is an integration constant.  

Now the Friedmann and Klein-Gordon equations can be rewritten as 
\begin{equation}
H^2 = \frac{\dot{\phi}^2}{2} + V + \frac{\rho_0}{f^{1-\gamma}a^{3(1+\gamma)}},
\label{Friedmann2}
\end{equation}
\begin{equation}
\ddot{\phi} + 3 H \dot{\phi} + V' + \frac{\gamma\rho_0 f'}{f^{1-\gamma}a^{3(1+\gamma)}} = 0.
\label{KG1} 
\end{equation}

We have chosen the cosmological evolution described by the Hubble variable (\ref{Hubble}). Correspondingly the cosmological radius $a(t)$ is 
\begin{equation}
a(t) = \frac{a_0 t^{\alpha}}{(t_R-t)^{\alpha}},
\label{radius}
\end{equation}
where $a_0$ is a positive constant. Now we can rewrite the matter energy density $\rho(t)$ as 
\begin{equation}
\rho(t) = \frac{M (t_R-t)^{3\alpha(1+\gamma)}}{f^{1-\gamma} t^{3\alpha(1+\gamma)}},
\label{matter1}
\end{equation}
where we have introduced a new constant
\begin{equation}
M = \frac{\rho_0}{a_0^{3(1+\gamma)}}.
\label{M}
\end{equation}
The term $\rho f$ in the right-hand side of the Friedmann equation (\ref{Friedmann2}) now looks as 
\begin{equation}
\rho f = \frac{M f^{\gamma}(t_R-t)^{3\alpha(1+\gamma)}}{ t^{3\alpha(1+\gamma)}}.
\label{matter2}
\end{equation}

Now, the left-hand side of Eq. (\ref{Friedmann2}) behaves as $\sim 1/t^2(t_R-t)^2$. If we would like the contribution of 
the kinetic term of the scalar field $\dot{\phi}^2/2$ to have the same structure we should require that the time dependence 
of the scalar field is 
\begin{equation}
\phi(t) = \phi_0\ln \frac{t}{t_R-t},
\label{scalar}
\end{equation}
where $\phi_0$ is a constant to be determined.  Then, we shall look for the time dependence of the potential $V$ in the form 
\begin{equation}
V = \frac{V_0 + V_1 t}{t^2(t_R-t)^2}. 
\label{potential}
\end{equation} 
We shall  try to find time dependence of the function $f$ from the relation 
\begin{equation}
f^{\gamma} = (f_0 + f_1 t) t^{3\alpha(1+\gamma) -2}(t_R-t)^{-3\alpha(1+\gamma) -2}.
\label{f}
\end{equation}
Substituting Eqs. (\ref{Hubble}), (\ref{scalar}), (\ref{potential}) and (\ref{f}) into Eq. (\ref{Friedmann2}) we 
obtain the following constraints on the constants :
\begin{equation}
\alpha^2 t_R^2 = M f_0 + \frac{\phi_0^2 t_R^2}{2} + V_0,
\label{constraint}
\end{equation}
\begin{equation}
V_1 + M f_1 = 0.
\label{constraint0}
\end{equation}

Now, let us consider the Klein-Gordon equation (\ref{KG1}). The second time derivative of the scalar field is  
\begin{equation}
\ddot{\phi} = -\frac{\phi_0}{t^2} + \frac{\phi_0}{(t_R-t)^2}= \frac{\phi_0t_R(2t-t_R)}{t^2(t_R-t)^2}.
\label{KG2}
\end{equation}
The ``friction'' term is 
\begin{equation}
3H\dot{\phi} = \frac{3\alpha\phi_0t_R^2}{t^2(t_R-t)^2}.
\label{KG3}
\end{equation}
The derivative of the potential is 
\begin{equation}
V'(\phi) = \frac{-2V_0t_R + (4V_0-V_1t_R)t + 3V_1 t^2}{\phi_0t_Rt^2(t_R-t)^2}.
\label{KG4}
\end{equation}
and the term, including the interaction with matter is 
\begin{eqnarray}
&&\gamma f' \rho = \frac{1}{\phi_0t_Rt^2(t_R-t)^2}\times(f_0t_R(3\alpha(1+\gamma)-2)\nonumber\\
&&+(4f_0+f_1t_R(3\alpha(1+\gamma)-1)+3f_1t^2).
\label{KG5}
\end{eqnarray}
Now, substituting the expressions (\ref{KG1})--(\ref{KG5}) into Eq. (\ref{KG1}) we obtain
the following constraints:
\begin{eqnarray}
&&-\phi_0^2t_R^2 + 3\phi_0^2t_R^2\alpha-2V_0 + Mf_0(3\alpha(1+\gamma)-2)=0, 
\label{constraint1}
\end{eqnarray}
\begin{eqnarray}
&&2\phi_0^2t_R^2+4V_0-V_1t_R + 4Mf_0 +Mf_1t_R(3\alpha(1+\gamma)-1) = 0,
\label{constraint2}
\end{eqnarray}
\begin{equation}
3V_1 + 3M f_1 = 0. 
\label{constraint3}
\end{equation}
The last condition (\ref{constraint3}) coincides with (\ref{constraint0}).

Now we can solve the constraints and find the relations between the parameters. We shall choose as free parameters 
the amplitude of the scalar field $\phi_0$ and the ``quantity of  matter '' $M > 0$.  
From Eqs. (\ref{constraint}) and (\ref{constraint2}) we find 
\begin{equation}
V_1 = \frac{4\alpha t_R}{3(1+\gamma)}.
\label{V1}
\end{equation}
Then from Eq. (\ref{constraint0}) follows 
\begin{equation}
f_1 = -\frac{4\alpha t_R}{3M(1+\gamma)}.
\label{f1} 
\end{equation}
From Eqs. (\ref{constraint}) and (\ref{constraint1}) we find 
\begin{equation}
f_0 = \frac{t_R^2(2\alpha - \phi_)^2)}{3M(1+\gamma)}
\label{f0}
\end{equation}
and 
\begin{equation}
V_0 = \frac{t_R^2(3\alpha(3\alpha(1+\gamma)-2)+3\phi_0^2(1-\gamma))}{6(1+\gamma)}.
\label{V0}
\end{equation}

We should also now express the time parameter $t$ in terms of the scalar field $\phi$.
It follows immediately from Eq. (\ref{scalar}) that 
\begin{equation}
t = \frac{t_R}{1 + \exp\left(-\frac{\phi}{\phi_0}\right)}.
\label{time-phi}
\end{equation}

Substituting the parameters $V_0, V_1, f_0$ and $f_1$ from Eqs. (\ref{V1})--({V0}) and the expression for the 
time parameter $t$ from Eq. (\ref{time-phi}) into Eqs. (\ref{potential}) and (\ref{f}) we find the explicit expressions 
for the potential $V(\phi)$ and the function $f(\phi)$:
\begin{eqnarray}
&&V(\phi) = \frac{8\cosh^4\frac{\phi}{2\phi_0}}{3(1+\gamma)}\left(6\alpha^2(1+\gamma)+3\phi_0^2(1-\gamma) + 4\alpha\tanh\frac{\phi}{2\phi_0}\right),
\label{potential1}
\end{eqnarray}
\begin{eqnarray}
&&f(\phi) = \left(-\frac{16\cosh^4\frac{\phi}{2\phi_0}\exp\left(3\alpha(1+\gamma)\frac{\phi}{\phi_0}\right)}
{3M t_R^2 (1+\gamma)}\left(3\phi_0^2+2\alpha\tanh\frac{\phi}{2\phi_0}\right)\right)^{\frac{1}{\gamma}}.
\label{f2}
\end{eqnarray}

\section{Parameters of the model}
The parameters $\alpha$ and $t_R$ are given by the cosmological dynamics of the model (\ref{Hubble}). 
The parameter $\phi_0$  is free. Without loosing generality we can choose it positive.
Correspondingly the  value of the scalar field $\phi$ (see Eq. (\ref{scalar}) )
runs from $-infty$ at the Big Bang beginning of the cosmological evolution to $+\infty$ at its end at the moment 
$t = t_R$ when it encounters the Big Rip singularity. It crosses the zero value at $t = t_R/2$, i.e. at the moment 
of the phantom divide line crossing.  Now, looking at the expression for the function $f(\phi)$ describing the 
interaction between the chameleon scalar field and the matter, we see that it is well defined only if 
the expression, which should be elevated into the power $1/\gamma$ is always positive, or if the parameter $\gamma$ satisfies 
some additional restrictions. Thus, we can consider two cases.\\

The case I. The expression in the bracket in Eq. (\ref{f2}) is always positive. It is possible if and only if 
\begin{equation}
\phi_0 \geq \sqrt{\frac{2\alpha}{3}},
\label{ineq}
\end{equation}
\begin{equation}
\gamma < -1.
\label{ineq1}
\end{equation}
Then if the parameter $\alpha > \frac13$ and 
\begin{equation}
\phi_0 > \sqrt{\frac{4\alpha - 6\alpha^2(1+\gamma)}{3(1-\gamma)}} 
\end{equation}
the potential $V(\phi)$ is always negative. If $\alpha > \frac13$ and 
\begin{equation}
\sqrt{\frac{2\alpha}{3}} < \phi_0 < \sqrt{\frac{4\alpha - 6\alpha^2(1+\gamma)}{3(1-\gamma)}} 
\end{equation}
the potential $V(\phi)$ changes  sign at 
\begin{equation}
\phi = 2\phi_0\  {\rm arctanh} \frac{5\alpha^2(1+\gamma) + 3\phi_0^2(1-\gamma)}{4\alpha}.
\end{equation}
If $ \alpha < \frac13$ the potential is always negative.

The case II. If at least one of two inequalities (\ref{ineq}), (\ref{ineq1}) is broken the expression for $f^{\gamma}$ in Eq. (\ref{f2}) cannot be always nonnegative. Hence, we should impose a following condition on the factor $\gamma$:
\begin{equation}
\gamma = \frac{2m+1}{n},
\label{gamma}
\end{equation}
where $m$ and $n$ are integers. In this case the expression for $f$ is well defined. 
The sign of the potential depends on the interplay of three parameters $\phi_0, \gamma$ and $\alpha$.

\section{Conclusion}
We have constructed an exact  solution for a particular Chameleon cosmological model of the type considered 
in paper \cite{Fara-Sale}. The universe in this solution begins its evolution from Big Bang singularity, 
undergoes a phantom divide line crossing and ends in the Big Rip singularity. The two potential-like functions 
of the chameleon scalar field have a rather  simple analytic form. 
Note that this form is simpler than the potential functions in two-scalar model, providing the same cosmological evolution
\cite{we-PT1}.  Thus, we hope that our version of the Chameleon model can be useful for the analysis of the 
phantom divide line crossing phenomenon. 

\section*{Acknowledgments} This work was partially supported by Grants RFBR  08-02-00923  and  LSS-4899.2008.2.


\begin{thebibliography}{99}
\bibitem{cosmic}
A.~G.~Riess {\it et al.}  [Supernova Search Team Collaboration],
  {\it Astron.\ J.}  {\bf 116} (1998) 1009;

S.~Perlmutter {\it et al.}  [Supernova Cosmology Project Collaboration],
  {\it Astrophys.\ J.}  {\bf 517} (1999) 565.

\bibitem{dark}
V.~Sahni and A.~A.~Starobinsky,
  {\it Int.\ J.\ Mod.\ Phys.\ D} {\bf 9} (2000) 373;

V.~Sahni and A.~A.~Starobinsky,
    {\it Int.\ J.\ Mod.\ Phys.\  D} {\bf 15} (2006) 2105.

T.~Padmanabhan,
   {\it Phys.\ Rept.}  {\bf 380} (2003) 235;

P.~J.~E.~Peebles and B.~Ratra,
  {\it Rev.\ Mod.\ Phys.}  {\bf 75} (2003) 559;

E.~J.~Copeland, M.~Sami and S.~Tsujikawa,
  {\it Int.\ J.\ Mod.\ Phys.\ D} {\bf 15} (2006) 1753.

\bibitem{phantom}
R.~R.~Caldwell,
  {\it Phys.\ Lett.\ B}   {\bf 545} (2002) 23.

\bibitem{Rip}
A.~A.~Starobinsky,
   {\it Grav.\ Cosmol.}  {\bf 6} (2000) 157;

R.~R.~Caldwell, M.~Kamionkowski and N.~N.~Weinberg,
 {\it Phys.\ Rev.\ Lett.}  {\bf 91} (2003) 071301.

\bibitem{observe}
U.~Alam, V.~Sahni, T.~D.~Saini and A.~A.~Starobinsky,
   {\it Mon.\ Not.\ Roy.\ Astron.\ Soc.}  {\bf 354} (2004) 275.

\bibitem{obs-cross}
V. Barger, Y. Gao  and D. Marfatia, {\it Phys. Lett.} B {\bf 648} (2007) 127;\\
A. Gong and A. Wang, {\it Phys. Rev.} D {\bf 75} (2007)  0435520;\\
U. Alam, V. Sahni  and A.A. Starobinsky, {\it JCAP} {\bf 0702} (2007)011;\\
S. Nesseris  and L. Perivolaropoulos, {\it JCAP} {\bf 0702} (2007) 025;\\
G. B. Zhao, J. Q. Xia, H. Li  et al,  {\it Phys. Lett.} B {\bf 648} (2007) 8;\\
P. Serra, A. Heavens  and A. Melchiorri, {\it MNRAS} {\bf 379} 1 (2007) 169;\\
T.M. Davis, E. Mortsell, J. Sollerman et al., 
{\it Astrophys. J.} {\bf 666} (2007) 716;\\
E. l. Wright,  
{\it Astrophys. J.} {\bf 664} (2007) 633;\\
Y. Wang  and P. Mukherjee, {\it Phys. Rev. } D {\bf 76} (2007) 103533.
\bibitem{divide}
B. Boisseau, G. Esposito-Farese, D. Polarski  and A.A. Starobinsky,
{\it Phys. Rev. Lett.} {\bf 85} (2000) 2236;\\
G. Esposito-Farese  and D. Polarski,  {\it Phys. Rev.} D {\bf 63} (2001) 063504;\\
A. Vikman,
{\it Phys. Rev.}  D {\bf 71} (2005) 023515;\\
L. Perivolaropoulos, 
{\it Phys.  Rev.} D {\bf 71} (2005)  063503;\\
B. McInnes, {\it Nucl. Phys.} B {\bf 718} (2005) 55;\\
I. Ya. Aref'eva, A.S. Koshelev  and S. Yu.Vernov,   {\it Phys. Rev.} D {\bf 72} (2005) 064017;\\
L. Perivolaroupoulos,   {\it JCAP} {\bf 0510} (2005) 001;\\
R.R. Caldwell and M. Doran,   {\it
Phys. Rev.} D {\bf 72} (2005) 043527;\\
V. Sahni and Yu. Shtanov,  {\it JCAP} {\bf 0311} (2003) 014;\\
V. Sahni  and L. Wang,  {\it Phys. Rev.} D {\bf 62} (2000)  103517;\\
H. Wei and R.G. Cai,  {\it Phys. Rev.} D {\bf 72} (2005) 123507;\\
H. Wei and R.G. Cai,  {\it Phys. Lett.} B {\bf 634} (2006)  9;\\
R.G. Cai  and A. Wang,  {\it JCAP} {\bf 0503} (2005) 002.


\bibitem{two-exp}

X.~L.~Feng, X.~L.~Wang and X.~M.~Zhang,
 {\it Phys.\ Lett.\  B} {\bf 607} (2005) 35;

L.~Perivolaropoulos,
{\it Phys.\ Rev.\  D} {\bf 71} (2005) 063503;

G.~B.~Zhao, J.~Q.~Xia, M.~Li, B.~Feng and X.~Zhang,
  {\it Phys.\ Rev.\  D} {\bf 72} 123515 (2005);

J.~Q.~Xia, B.~Feng and X.~M.~Zhang,
  {\it Mod.\ Phys.\ Lett.\  A} {\bf 20} 2409 (2005);

B.~Feng, M.~Li, Y.~S.~Piao and X.~Zhang,
  {\it Phys.\ Lett.\  B} {\bf 634} 101 (2006);

Z.~K.~Guo, Y.~S.~Piao, X.~M.~Zhang and Y.~Z.~Zhang,
  {\it Phys.\ Lett.\  B} {\bf 608} 177 (2005);

H.~Wei, R.~G.~Cai and D.~F.~Zeng,
  {\it Class.\ Quant.\ Grav.}  {\bf 22} 3189 (2005);

R.~Lazkoz and G.~Leon,
  {\it Phys.\ Lett.\  B} {\bf 638} 303 (2006);

X.~F.~Zhang and T.~Qiu,
  {\it Phys.\ Lett.\  B} {\bf 642} 187 (2006);

W.~Zhao,
  {\it Phys.\ Rev.\  D} {\bf 73} 123509 (2006);

M.~Alimohammadi and H.~M.~Sadjadi,
  {\it Phys.\ Lett.\  B} {\bf 648} 113 (2007);

Z.~K.~Guo, Y.~S.~Piao, X.~Zhang and Y.~Z.~Zhang,
  {\it Phys.\ Rev.\  D} {\bf 74} 127304 (2006);

Y.~f.~Cai, H.~Li, Y.~S.~Piao and X.~m.~Zhang,
  {\it Phys.\ Lett.\  B} {\bf 646} 141 (2007);

X.~Zhang,
  {\it Phys.\ Rev.\  D} {\bf 74} 103505 (2006);

M.~R.~Setare,
  {\it Phys.\ Lett.\  B} {\bf 641} 130 (2006);

Y.~f.~Cai, M.~z.~Li, J.~X.~Lu, Y.~S.~Piao, T.~t.~Qiu and X.~m.~Zhang,
  {\it Phys.\ Lett.\  B} {\bf 651} 1 (2007);

R.~Lazkoz, G.~Leon and I.~Quiros,
  {\it Phys.\ Lett.\  B} {\bf 649} 103 (2007);

M.~Alimohammadi,
  {\it Gen.\ Rel.\ Grav.}  {\bf 40} 107 (2008);

M.~R.~Setare and E.~N.~Saridakis,
  {\it Phys.\ Lett.\  B} {\bf 668} 177 (2008);

J.~Sadeghi, M.~R.~Setare, A.~Banijamali and F.~Milani,
  {\it Phys.\ Lett.\  B} {\bf 662} 92 (2008);

H.~H.~Xiong, Y.~F.~Cai, T.~Qiu, Y.~S.~Piao and X.~Zhang,
  {\it Phys.\ Lett.\  B} {\bf 666} 212 (2008);

Y.~F.~Cai and J.~Wang,
  {\it Class.\ Quant.\ Grav.}  {\bf 25} 165014 (2008);

S.~Zhang and B.~Chen,
  {\it Phys.\ Lett.\  B} {\bf 669} 4 (2008);

M.~R.~Setare and E.~N.~Saridakis,
  {Int.\ J.\ Mod.\ Phys.\  D} {\bf 18} 549 (2009);

M.~R.~Setare and E.~N.~Saridakis,
  {\it JCAP} {\bf 0809} 026 (2008);

M.~R.~Setare and E.~N.~Saridakis,
  {\it Phys.\ Lett.\  B} {\bf 671} 331 (2009);

K.~Nozari, M.~R.~Setare, T.~Azizi and N.~Behrouz,
  {\it Phys.\ Scripta} {\bf 80} 025901 (2009);

M.~R.~Setare and E.~N.~Saridakis,
  {\it Phys.\ Rev.\  D} {\bf 79} 043005 (2009);

L.~P.~Chimento, M.~Forte, R.~Lazkoz and M.~G.~Richarte,
  {\it Phys.\ Rev.\  D} {\bf 79} 043502 (2009);

E.~N.~Saridakis,
  arXiv:0903.3840 [astro-ph.CO];

H.~Wei and S.~N.~Zhang,
  {\it Phys.\ Rev.\  D} {\bf 76} 063005 (2007);

Y.~F.~Cai, E.~N.~Saridakis, M.~R.~Setare and J.~Q.~Xia,
  arXiv:0909.2776 [hep-th];

H.~Zhang,
  arXiv:0909.3013 [astro-ph.CO].

\bibitem{we-PT}
A.~A.~Andrianov, F.~Cannata and A.~Y.~Kamenshchik,
  {\it J.\ Phys.\ A}  {\bf 39} (2006) 9975;

\bibitem{we-PT1}
A.~A.~Andrianov, F.~Cannata and A.~Y.~Kamenshchik,
    {\it Int.\ J.\ Mod.\ Phys.\  D} {\bf 15} (2006) 1299.

\bibitem{we-PT2}
A.~A.~Andrianov, F.~Cannata, A.~Y.~Kamenshchik and D. Regoli,
    {\it Int.\ J.\ Mod.\ Phys.\  D} {\bf 19} (2010) 97.

\bibitem{we}
A.~A.~Andrianov, F.~Cannata and A.~Y.~Kamenshchik,
  {\it Phys.\ Rev.\  D} {\bf 72} (2005) 043531;


F.~Cannata and A.~Y.~Kamenshchik,
   {\it Int.\ J.\ Mod.\ Phys.\  D} {\bf 16} (2007) 1683;

\bibitem{non-minimal}
R.~Gannouji, D.~Polarski, A.~Ranquet and A.~A.~Starobinsky,
   {\it JCAP} {\bf 0609} (2006) 016.

\bibitem{we-two}
A.~A.~Andrianov, F.~Cannata, A.~Y.~Kamenshchik and D. Regoli,
    {\it JCAP} {\bf 0802} (2008) 015.

\bibitem{chameleon}
 J.~Khoury and A.~Weltman,
{\it  Phys.\ Rev.\ Lett.}  {\bf 93} (2004) 171104;

J.~Khoury and A.~Weltman,
  {\it Phys.\ Rev.\  D} {\bf 69} (2004) 044026;

P.~Brax, C.~van de Bruck, A.~C.~Davis, J.~Khoury and A.~Weltman,
  {\it Phys.\ Rev.\  D} {\bf 70} (2004) 123518.

\bibitem{Fara-Sale}
H. Farajollahi and A. Salehi, 
Cosmic Dynamics in the Chameleon Cosmology, arXiv:1004.3508 [gr-qc].  
\bibitem{power-law}
F.~Lucchin and S.~Matarrese,
  {\it Phys.\ Rev.\ D} {\bf 32} (1985) 1316;

J.~J.~Halliwell,
 {\it Phys.\ Lett.\  B} {\bf 185} (1987) 341.

A.~B.~Burd and J.~D.~Barrow,
{\it Nucl.\ Phys.\ B} {\bf 308} (1988) 929;

V.~Gorini, A.~Y.~Kamenshchik, U.~Moschella and V.~Pasquier,
 {\it Phys.\ Rev.\ D} {\bf 69} (2004) 123512.
 



\end{thebibliography}
\end{document}